\documentclass[prb,aps,twocolumn,amsmath,amssymb,floatfix,
superscriptaddress]{revtex4}

\usepackage{graphics,epstopdf}
\usepackage{epsf,graphics,graphicx}
\usepackage{color}

\date{\today}

\begin{document}

\title{Spin-selective transmission through a single-stranded magnetic helix}

\author{Suparna Sarkar}

\affiliation{Physics and Applied Mathematics Unit, Indian Statistical
Institute, 203 Barrackpore Trunk Road, Kolkata-700 108, India}

\author{Santanu K. Maiti}

\email{santanu.maiti@isical.ac.in}

\affiliation{Physics and Applied Mathematics Unit, Indian Statistical
Institute, 203 Barrackpore Trunk Road, Kolkata-700 108, India}

\begin{abstract}

Magnetic helix (MH) structure can be a role model for future spintronic 
devices. Utilizing the advantage of constructing possible magnetic 
configurations, in the present work first time we investigate spintronic 
behavior, to the best of our knowledge, in a helical geometry with finite 
magnetic ordering. The interplay between short-range and long-range hopping 
of electrons yields many non-trivial features which are thoroughly studied. 
Quite interestingly we see that the MH exhibits the strong chiral-induced 
spin selectivity effect, like what is observed in chiral molecules. Finally,
to make the model more realistic we also examine the effect of helical
dynamics. All the results are valid for a wide range of physical parameters, 
which prove the robustness of our analysis. 

\end{abstract}

\maketitle

\section{Introduction}

Proper development of reliable methods to manipulate spin of an electron
rather than its charge has been the subject of intense research in last 
few decades~\cite{wolf}. One of the key reasons is that electron's spin 
allows us to do more works providing much less effort, compared to the 
conventional electronic devices that are usually charge based. Remarkable 
progress has already been made in downsizing the functional elements, and 
considering the versatility of spintronic devices~\cite{s1,s2,s3,s4}, 
which are usually faster, efficient and smaller in sizes than conventional 
electronic systems, we may think that within few years all these electronic 
devices might be replaced by spintronic ones~\cite{s1,s2,s3,s4,s5,s6}.

Now in order to substantiate these facts the first and foremost thing is 
the generation of polarized spin current and its proper regulation. The 
use of ferromagnetic materials is one of the possible routes for 
it~\cite{xie,wang}. But whenever we need to tune spin current, generated 
by a ferromagnetic material, we think about the application of magnetic 
field though it is usually not so suitable especially for the small regions 
because of the difficulty in confining the magnetic field. To circumvent 
this issue, people were trying to design a spin filter using intrinsic 
properties~\cite{int1,int2,int3} of the materials, for instance, spin-orbit 
(SO) couplings~\cite{rashba,dressel,winkler}. Two types of SO couplings, 
namely, Rashba~\cite{rashba} and Dresselhaus~\cite{dressel}, are commonly 
used out of which Rashba SO coupling draws significant attention, specially 
due to the fact that it can be tuned {\em externally}~\cite{ex1,ex2,ex3,ex4} 
which yields selective spin transmission, while the other one cannot be 
regulated as it is material dependent. Several proposals were put 
forward~\cite{pr1,pr2,pr3,pr4,pr5,pr6,pr7,pr8} to get a polarized beam from a 
completely unpolarized one, employing the role of SO couplings, considering 
different tailor made geometries, molecular systems, etc. 
Among them, G\"{o}hler {\em et al.} have shown in their 
experimental work~\cite{ghl} that a very high degree of spin polarization 
can be achieved even at room temperature with the help of self-assembled 
monolayers of double-stranded DNA (dsDNA) molecules deposited on gold 
substrate. A light is incident on the substrate which emits photoelectrons 
with both the two spin components, and when they get transmitted through 
the other end of the dsDNA, they become highly polarized. 
This experiment~\cite{ghl} essentially opens up a new possibility to design 
efficient spin filters using helical molecules, and the phenomenon is 
referred as chiral-induced spin selectivity 
(CISS)~\cite{ciss1,ciss2,ciss3,ciss4,ciss5}.

Soon after this experiment, enormous attention has been paid in analyzing 
CISS effect both experimentally as well as theoretically considering 
different chiral systems~\cite{ciss1,ciss2,ciss3,ciss4,ciss5,qfs1}. 
So far, to the best of our our concern, three 
research groups, viz, Guti\'{e}rrez {\em et al.}~\cite{ciss4}, Medina and 
co-workers~\cite{ciss5} and Guo {\em et al.}~\cite{qfs1} 
have essentially explored theoretically the phenomenon of CISS effect in 
helix-shaped geometries and DNA molecules. Considering a simple helical 
geometry, instead of using a DNA molecule or $\alpha$-helical peptide,
Guti\'{e}rrez {\em et al.} have shown that reasonably large spin polarization 
can be achieved near the energy band edges for a wide parameter range of
electron coupling and SO interaction. For this geometry the spin polarization 
is obtained due to the interaction of the spin with the magnetic field, 
and this field is generated as a result of the motion of electrons in 
helical electrostatic
potential~\cite{ciss4}. In another theoretical prescription, Medina and 
co-workers have shown~\cite{ciss5} spin polarization in a simple chiral 
molecule composed of six carbon atoms. Here also the SO coupling plays the 
central role for getting spin selectivity, though the degree of spin 
polarization is quite less as observed in DNA experiments. It has been 
claimed that the spin polarization efficiency can be enhanced with increasing 
the density of carbon atoms~\cite{ciss5}. Lastly, in the other proposal made
by Guo and Sun, it has been shown that CISS effect can be observed in a 
ds-DNA molecule. Their work has been based on a special ansatz that along
with Rashba SO coupling and helical symmetry, dissipation effect is also
required. If any one among these factors is absent, then no spin polarization 
will be visible, as suggested by Guo and Sun~\cite{qfs1}, though the 
underlying physical mechanism behind this is not fully clear. Although 
the above three prescriptions are quite different, they have a common 
signature which is the chirality of the geometry.

Very recently another proposition was made where CISS effect has been 
predicted in {\em single stranded} helical molecule with longer range hopping 
integrals~\cite{qfs2,qfs3}. It is the protein-like $\alpha$-helical molecule. 
It has been suggested that while a single-stranded DNA (ssDNA) is too poor 
in spin selectivity, the $\alpha$-helical protein on the other hand provides 
reasonably large spin polarization~\cite{qfs2}. The basic mechanism is hidden 
within the hopping of electrons, in one case (ssDNA) it is restricted within 
almost nearest-neighbors, while for the protein molecule longer-range hoppings 
are allowed that lead to the significant difference between these look-wise 
identical molecular systems. Thus, chiral system with longer-range hopping 
integrals might be a role model for designing efficient spintronic 
devices~\cite{qfs2,qfs3}.

For all the above mentioned studies the key factor that is 
involved to have CISS effect is the SO coupling. Now, it is well-known that 
the SO coupling is too weak for the helical molecular systems~\cite{smSO}, 
especially for DNA ones, and the other fact is that for all these cases the
polarization is achieved only at energy resonances. Thus, a questions 
naturally arises that can we think about any other alternative helical 
geometry that may exhibit favorable spin separation in presence of any 
other kind of spin-dependent scattering, and useful for CISS effect.
Mimicking the structure of single-stranded helical biological 
molecules, in the present work we consider such a tailor made magnetic helical 
geometry, called as magnetic helix (MH), to explore spin selective electron 
transmission and CISS effect. Considering a MH structure, no analysis has 
been made so far along these lines, to the best of our knowledge. We explore a
strong CISS effect, analogous to previous analysis, and most importantly, a 
high degree of spin polarization is obtained for a wide range of physical
parameters. Here it is important to note that, we do not need to consider 
a double-stranded system as well as environmental dephasing. The interplay 
between short-range and long-range hopping of electrons is also critically 
discussed, and we find several interesting patterns in spin selectivity. 
In addition to spin separation, it is always beneficial if we can tune spin 
selectivity by some external means. We do it with the help of gate voltage 
by placing the functional element, MH, within the suitable gate electrodes. 
The gate controlled transport properties in different geometrical structures 
have been reported in several other contemporary works~\cite{gc1,gc2,gc3,
gc4,gc5}.
Finally, to make the model more realistic we include the 
effect of helical dynamics. It has been clearly explained by Guti\'{e}rrez
{\em et al.}~\cite{hd1}, Ratner {\em et al.}~\cite{hd2,hd3,hd4} and some 
other groups~\cite{hd5,hd6,hd7} that the static 
picture do not give the complete scenario for physical phenomena, and thus, 
one needs to go beyond that i.e., dynamics of the system needs to be 
incorporated for the sake of completeness.

The remaining part of the work is arranged as follows. The model of magnetic 
helix along with the theoretical prescription for the calculations of spin
dependent transport are described in Sec. II. All the essentials results are
critically explained in Sec. III, and finally in Sec. IV we conclude our 
findings.

\section{Magnetic helix, TB Hamiltonian and theoretical framework}

\subsection{Magnetic helix and the TB Hamiltonian}

\begin{figure}[ht]
{\centering \resizebox*{5.5cm}{6.5cm}{\includegraphics{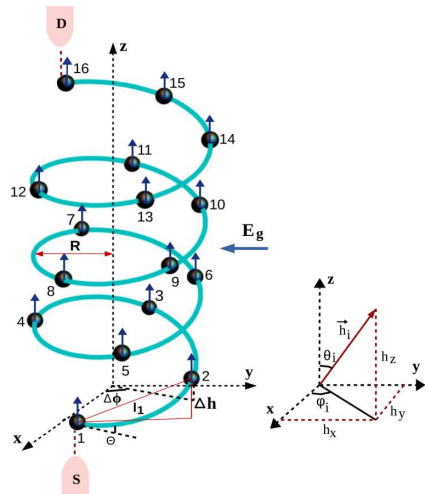}}\par}
\caption{(Color online). Left panel: schematic diagram of a right-handed 
helical geometry, coupled to source and drain electrodes, having radius $R$, 
stacking distance $\Delta h$, twisting angle $\Delta \phi$, helix angle 
$\Theta$ and nearest-neighbor distance $l_1$. Each site $i$ of the system 
contains a finite magnetic moment which is responsible for spin dependent 
scattering through the interaction of the injected electron spin with the 
net spin $\langle \vec{S}_i\rangle$ located at $i$th site. We define this 
scattering parameter as $\vec{h}_i$. An external electric 
field $E_g$ is applied to the system, along the perpendicular direction of 
the helix axis, by placing it within suitable gate electrodes. Right panel: 
general arrangement of $\vec{h}_i$ associated with the net spin at $i$th
site, where $\theta_i$ and $\varphi_i$ are the polar and azimuthal angles, 
respectively.}
\label{model}
\end{figure}

Let us begin with the junction setup shown in the left panel of 
Fig.~\ref{model}, where a right-handed magnetic helix is coupled to source 
(S) and drain (D) electrodes. 
Each site of the helix, labeled as 1, 2, 3 $\ldots$, is subjected to a
finite magnetic moment. The incoming electron interacts with these local
magnetic sites through the usual spin-spin exchange coupling $J$.
If $\langle \vec{S}_i \rangle$ be the net spin at the magnetic site $i$, 
then we can define the spin-dependent scattering parameter at $i$th site as 
$\vec{h}_i=J\langle \vec{S}_i \rangle$~\cite{spsp}. The orientation of 
$\vec{h}_i$,
associated with the orientation of $\langle \vec{S}_i\rangle$ is described 
by the polar angle $\theta_i$ and azimuthal angle $\varphi_i$ (right panel), 
as used in conventional spherical polar co-ordinate system. These magnetic 
sites are responsible for spin dependent scattering. The parameters 
$\Delta h$ and $\Delta \phi$, describing the stacking distance and twisting 
angle respectively, play important roles to characterize whether the 
long-range or short-range hopping is important, and at the same time 
determines the structure of the magnetic helix. When $\Delta h$ is too 
small i.e., the atoms are closely spaced, electrons can hop to different 
sites of the geometry yielding a long-range hopping (LRH) helix, while 
for the other situation when $\Delta h$ is reasonably large the system 
maps to the short-range hopping (SRH) helix. As already stated, depending 
on $\Delta h$ and $\Delta \phi$, the hopping of electrons as well as the 
helical structure get modified which can be more clearly understood from 
our forthcoming analysis. We will examine the effects of all these factors 
on spin selective transmission in detail for the comprehensive analysis.

In order to describe the Hamiltonian of the magnetic helix sandwiched 
between S and D (left panel of Fig.~\ref{model}), a tight-binding (TB) 
framework is given. In presence of 
magnetic interaction~\cite{mh1,mh2,mh3,mh4}, the TB Hamiltonian of a 
right-handed magnetic helix having $N$ sites reads as, 
\begin{eqnarray}
H_{\mbox{\tiny mh}} & = & \sum_i \mbox{\boldmath{$c$}}_i^{\dagger} 
\left(\mbox{\boldmath{$\epsilon$}}_i-\mbox{\boldmath{$h$}}_i.\mbox{\boldmath{$\sigma$}}\right)\mbox{\boldmath{$c$}}_i
\nonumber \\
 & + & 
\sum_{i=1}^{N-1}\sum_{j=1}^{N-i}\left(\mbox{\boldmath{$c$}}_i^{\dagger}\mbox{\boldmath{$t$}}_j \mbox{\boldmath{$c$}}_{i+j} + 
\mbox{\boldmath{$c$}}_{i+j}^{\dagger}\mbox{\boldmath{$t$}}_j^{\dagger}\mbox{\boldmath{$c$}}_{i}\right)
\label{equ2}
\end{eqnarray}
where,
$\mbox{\boldmath{$c$}}_i=\begin{pmatrix}
c_{i\uparrow} \\
c_{i\downarrow} 
\end{pmatrix}$,
$\mbox{\boldmath{$c$}}_i^{\dagger}=\begin{pmatrix}
c_{i\uparrow}^{\dagger} & c_{i\downarrow}^{\dagger} 
\end{pmatrix}$,
$\mbox{\boldmath{$\epsilon$}}_i=\begin{pmatrix}
\epsilon_i & 0\\
0 & \epsilon_i
\end{pmatrix}$,
$\mbox{\boldmath{$t$}}_j=\begin{pmatrix}
t_j & 0\\
0 & t_j
\end{pmatrix}$,
and $\mbox{\boldmath{$h$}}_i.\mbox{\boldmath{$\sigma$}}=h_i\begin{pmatrix}
\cos\theta_i & \sin\theta_i e^{-k\varphi_i}\\
\sin\theta_i e^{k\varphi_i} & -\cos\theta_i
\end{pmatrix}$.
The term ($\mbox{\boldmath{$\epsilon$}}_i-\mbox{\boldmath{$h$}}_i.\boldsymbol{\sigma}$) represents the effective site energy matrix, where
$\epsilon_i$ is the on-site energy in absence of any magnetic interaction,
and $\mbox{\boldmath{$h$}}_i.\mbox{\boldmath{$\sigma$}}$ is the spin dependent 
scattering factor which appears due 
to the presence of magnetic moments at each lattice sites of the geometry. 
$\mbox{\boldmath{$\sigma$}}=(\sigma_x, \sigma_y, \sigma_z)$ is the Pauli
spin vector with the component $\sigma_z$ in diagonal representation.
$c_{i\sigma}^{\dagger}$ and $c_{i\sigma}$ are the usual fermionic creation 
and annihilation operators, respectively, at $i$th site of spin $\sigma$
($\uparrow, \downarrow$). The meanings of $\theta_i$ and $\varphi_i$ are
already described above, and here $k$ corresponds to $\sqrt{-1}$. The 
parameter $t_j$ is associated with the electron hopping between the sites
$i$ and ($i+j$), and it is expressed as~\cite{qfs3}
$t_j=t_1 e^{-(l_j-l_1)/l_c}$, where $t_1$ gives the nearest-neighbor 
hopping (NNH) integral, $l_1$
represents the nearest-neighbor distance (see Fig.~\ref{model}), and $l_c$
denotes the decay constant. The distance $l_j$ between the sites $i$ and 
($i+j$) is written in terms of the radius $R$, twisting angle $\Delta \phi$, 
and stacking distance $\Delta h$ as~\cite{qfs3} 
$l_j=\sqrt{[2R\sin(j\Delta\phi/2)]^2+(j\Delta h)^2}$.

Now, in presence of an external electric field $E_g$, perpendicular to the 
helix axis, the site energy of the helical geometry gets modified~\cite{efl} 
and it becomes 
\begin{equation}
\epsilon_i = \epsilon_i^0 + e V_g \cos\left(i\Delta \phi-\beta\right)
\label{equ5}
\end{equation}
where $\epsilon_i^0$ is the site energy in absence of $E_g$, and $V_g$ is
the gate voltage which is related to $E_g$ as $2V_g=2E_g R$ ($R$ being the
radius, see Fig.~\ref{model}). The phase factor
$\beta$ is associated with the electric field direction. It measures the
angle between the incident electric field and the positive $X$ axis. This
phase can be changed quite easily either by rotating the helical geometry or
by changing the positions of the gate electrodes (for schematic illustration
of this kind of electric field, see Ref.~\cite{efl}).

This is all about the TB Hamiltonian of the magnetic helix. Now we describe
the TB Hamiltonians of the other parts of the junction i.e., the contact 
electrodes S and D and their coupling with the MH. The electrodes are 
assumed to be perfect, reflection-less, non-magnetic and semi-infinite. 
Considering the site energy and NNH integrals (long-range hopping is not 
considered in the electrodes) as $\epsilon_0$ and $t_0$, respectively, we 
can express the TB Hamiltonians as
\begin{equation}
H_{\mbox{\tiny S}} =\sum_{m<1}\mbox{\boldmath{$a$}}_m^{\dagger} 
\mbox{\boldmath{$\epsilon$}}_0\mbox{\boldmath{$a$}}_m +
\sum_{m<1} (\mbox{\boldmath{$a$}}_m^{\dagger} \mbox{\boldmath{$t$}}_0 
\mbox{\boldmath{$a$}}_{m-1}+h.c.)
\end{equation}
and
\begin{equation}
H_{\mbox{\tiny D}} = \sum_{m>N} \mbox{\boldmath{$b$}}_m^{\dagger} 
\mbox{\boldmath{$\epsilon$}}_0 \mbox{\boldmath{$b$}}_m +
\sum_{m>N} (\mbox{\boldmath{$b$}}_m^{\dagger} \mbox{\boldmath{$t$}}_0 
\mbox{\boldmath{$b$}}_{m+1}+h.c.)
\end{equation}
where $\mbox{\boldmath{$\epsilon$}}_0$ and $\mbox{\boldmath{$t$}}_0$ are the 
($2\times 2$) diagonal matrices, and the forms of $\mbox{\boldmath{$a$}}_m$ and 
$\mbox{\boldmath{$b$}}_m$ are similar to what is described above for 
$\mbox{\boldmath{$c$}}_i$. The operators ($a_{m\sigma}^{\dagger}$, $a_{m\sigma}$) 
and ($b_{m\sigma}^{\dagger}$, $b_{m\sigma}$) are the usual fermionic operators
used for source and drain electrodes.

These electrodes are directly coupled at the two extreme points of the 
helical geometry as shown in Fig.~\ref{model}. In terms of the coupling 
strengths $t_S$ and $t_D$ due to source and drain, respectively, the TB 
coupling Hamiltonian looks like 
\begin{equation}
H_{\mbox{\tiny cpl}} = \mbox{\boldmath{$a$}}_0^\dagger
\mbox{\boldmath{$t$}}_S \mbox{\boldmath{$c$}}_1 +
\mbox{\boldmath{$c$}}_N^\dagger \mbox{\boldmath{$t$}}_D
\mbox{\boldmath{$b$}}_{N+1}+h.c.
\label{equ4}
\end{equation}
where $\mbox{\boldmath{$t$}}_S$ and $\mbox{\boldmath{$t$}}_D$ 
are the ($2\times 2$) diagonal matrices.

\subsection{Theoretical prescription}

To compute spin dependent transport quantities we use Green's function 
formalism, a standard and suitable technique for such calculations~\cite{gf1}. 
The fundamental quantity that is first required to determine is the 
two-terminal transmission function. In terms of coupling matrices 
$\mbox{\boldmath{$\Gamma$}}_S^{\sigma}$ and 
$\mbox{\boldmath{$\Gamma$}}_D^{\sigma^{\prime}}$, the transmission function 
becomes~\cite{gf1,gf2,gf3}
\begin{equation}
T_{\sigma\sigma^\prime} = \mbox{Tr}\left[\mbox{\boldmath{$\Gamma$}}_S^\sigma \mbox{\boldmath{$G$}}^r 
\mbox{\boldmath{$\Gamma$}}_D^{\sigma^\prime} \mbox{\boldmath{$G$}}^a\right]
\label{equ6}
\end{equation}
where $\mbox{\boldmath{$\Gamma$}}_{S(D)}^{\sigma(\sigma^\prime)}=-2 \mbox{Im}
\left[\mbox{\boldmath{$\Sigma$}}_{S(D)}^{\sigma(\sigma^\prime)}\right]$.
$\mbox{\boldmath{$\Sigma$}}_{S(D)}^{\sigma}$ is the contact self-energy due to source (drain)
electrode, which incorporates the effect of contact electrode. The factors
$\mbox{\boldmath{$G$}}^r$ and $\mbox{\boldmath{$G$}}^a$ are the retarded 
and advanced Green's functions, respectively, and they are determined via 
the relations~\cite{gf1}
$\mbox{\boldmath{$G$}}^r=(\mbox{\boldmath{$G$}}^a)^{\dagger} = 
\left[E \mbox{\boldmath{$I$}} - H_{\mbox{\tiny mh}} - \boldsymbol{\Sigma}_S^\sigma 
-\boldsymbol{\Sigma}_D^\sigma\right]^{-1}$, where $\mbox{\boldmath{$I$}}$ 
is the identity matrix having dimension ($2N\times 2N$). $T_{\sigma\sigma^\prime}$
describes the probability of a transmitted electron with spin 
$\sigma^{\prime}$ that is injected with spin $\sigma$. Thus, for 
$\sigma=\sigma^{\prime}$, we get pure spin transmission, while for the
other case ($\sigma \ne \sigma^{\prime}$) spin-flip transmission is obtained.
We define the net up and down spin transmission probabilities as:
$T_{\uparrow}=T_{\uparrow\uparrow}+T_{\downarrow\uparrow}$
and $T_{\downarrow}=T_{\downarrow\downarrow}+T_{\uparrow\downarrow}$.

Determining spin dependent transmission probabilities, we can easily 
calculate different spin dependent currents by integration procedure of the
transmission function over a suitable energy window associated with
bias voltage $V$. At zero temperature, the current expression 
becomes~\cite{gf1}
\begin{equation}
I_{\sigma\sigma^\prime} = \displaystyle \frac{e}{h} \int 
\limits_{E_F-\frac{eV}{2}}^{E_F+\frac{eV}{2}} T_{\sigma\sigma^\prime}(E)\,dE
\label{equ7}
\end{equation}
where $e$ and $h$ are the fundamental constants, and $E_F$ is the equilibrium
Fermi energy. Since the broadening of energy levels due to the coupling of
the MH with the contact electrodes is too large compared to the thermal 
broadening~\cite{gf1}, we can safely ignore the effect of temperature in 
our present analysis.

Using spin dependent currents, we eventually compute spin polarization 
coefficient from the relation~\cite{mh4,pola1}
\begin{equation}
P=\frac{I_\uparrow-I_\downarrow}{I_\uparrow+I_\downarrow}
\label{equ8}
\end{equation}
where $I_{\uparrow}=I_{\uparrow\uparrow}+I_{\downarrow\uparrow}$
and $I_{\downarrow}=I_{\downarrow\downarrow}+I_{\uparrow\downarrow}$.
$P=0$ means no polarization, whereas $P=\pm 1$ corresponds to $100\%$ 
spin polarization.

Finally, to include the effect of helical dynamics within 
the framework of TB approximation, we follow the prescription as originally
put forward by Ratner and co-workers~\cite{hd2}, an elegant and simple idea 
to incorporate the dynamics of a system. The site energy of a particular site
of the magnetic helix, where the drain electrode is coupled, gets modified
by adding an imaginary part $-k\hbar/\tau$ (i.e., $\epsilon_i \rightarrow
\epsilon_i -k\hbar/\tau$, $k=\sqrt{-1}$). The parameter $\tau$ represents the
decay time, and its value should be chosen in such a way that the electron 
vanishes immediately when it will reach to the drain end, and at the same 
time emphasis should be given that it cannot be so small that wave function 
gets reflected from the end site~\cite{hd2}.

\section{Numerical Results and discussion}

\begin{figure}[ht]
{\centering \resizebox*{8cm}{9cm}{\includegraphics{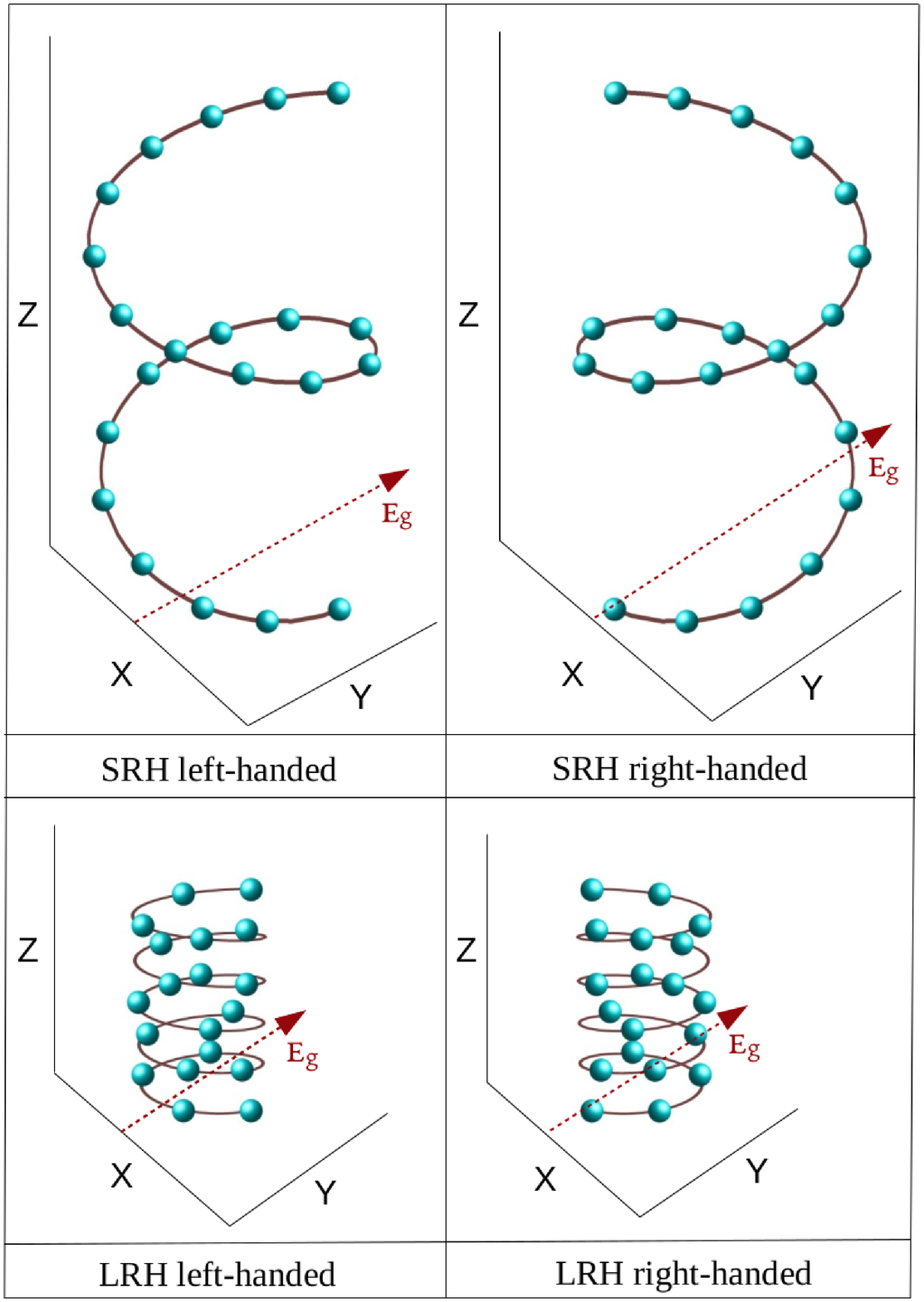}}\par}
\caption{(Color online). Both right-handed and left-handed magnetic helices
under two different hopping conditions.} 
\label{allmh}
\end{figure}

In what follows we present our essential results of spin dependent transport. 
The key aspects that we want to explore are: (i) the interplay between SRH 
and LRH interactions on transport phenomena, (ii) the CISS effect and (iii) 
the helical dynamics on spin selective transmission.
Unless stated, we select the physical parameters in such a way that we can 
have two different magnetic helix systems: one associated with the long-range
hopping and the other associated with the short-range one.
If not specified, we choose the parameters for the right-handed LRH MH as
$R=2.5\,$\AA, $\Delta h=1.5$\AA~and $\Delta \phi= 5\pi/9$, while these
parameters for the right-handed SRH MH are: $R=7$\AA, $\Delta h=3.4$\AA~and 
$\Delta \phi= \pi/5$. The decay exponent is fixed as $l_c=0.9$\AA.
These parameters are equivalent to the DNA and $\alpha$-helical protein 
molecules, the suitable examples of SRH and LRH models as already
established in literature~\cite{rps}. 
Along with the interplay between two kinds of hopping of electrons, in 
the present work we also consider both the right- and left-handed
magnetic helices to explore the strong CISS effect. As it is already stated
earlier that depending on $\Delta h$ and $\Delta \phi$, the structure of
the MH also gets changed together with the hopping. To reveal this fact
look into the spectra given in Fig.~\ref{allmh}.
What we see is that, for the chosen set of parameter values, the short-range
hopping takes place along the MH with about two turn when $N=21$, while about 
five turns occurs in the case of long-range hopping if we set $N=19$. It 
clearly demonstrates the structural change of the helical geometry, and we 
explore these issues in spin selective electron transmission. 

The other common parameters that we  use throughout the analysis are as 
follows. The on-site energy and NNH integrals,
$\epsilon_0$ and $t_0$, in the contact electrodes are fixed at zero and 
$4$ respectively, and these electrodes are connected to the MH with the 
strength $t_S=t_D=0.8$. The field-independent site energy $\epsilon_i^0$ in 
the MH is set at zero, without loss of any generality, and we fix $t_1=1$.
Unless specified, we fix the system size $N=20$. For the sake of
simplification, we set the polar and azimuthal angles to zero, and 
under this condition no spin-flip transmission is available. 
In our numerical calculations, the left-handed helix is obtained by 
changing $\Delta \phi$ to $-\Delta \phi$~\cite{gc5}, and if not mentioned
we present the results of right-handed magnetic helices. Finally,
we consider the magnitude of spin-dependent interaction parameter 
$h_i=1\,$eV $\forall\,i$. All the other energies are also measured in unit 
of electron volt (eV). Here it is important to note that the magnitude of
$\vec{h}_i$ ($=J\langle \vec{S}_i \rangle$) cab be even much higher than
$1\,$eV due to the strong coupling~\cite{spsp} (defined by the strength 
$J$) between 
the spin of an incoming electron with the net spin $\langle \vec{S}_i\rangle$
of the local magnetic sites in the magnetic helix. This is one of the key
advantage of getting strong spin-dependent scattering in a magnetic material
compared to the spin-orbit coupled systems~\cite{spsp}.

\subsection{Interplay between SRH and LRH interactions on transport phenomena}

\begin{figure}[ht]
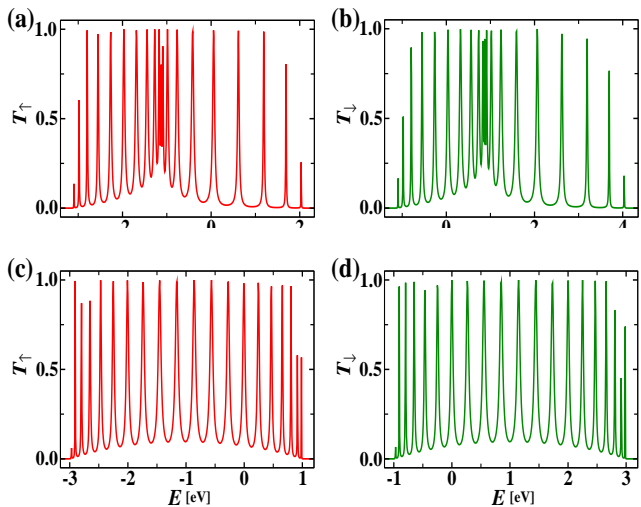

{\centering\resizebox*{4.2cm}{3.3cm}{\includegraphics{protranfig1.eps}}
\resizebox*{4.2cm}{3.3cm}{\includegraphics{protranfig2.eps}}
\resizebox*{4.2cm}{3.3cm}{\includegraphics{dnatranfig1.eps}}
\resizebox*{4.2cm}{3.3cm}{\includegraphics{dnatranfig2.eps}}\par}
\caption{(Color online). Up and down spin transmission probabilities as a
function of injecting electron energy $E$ for a MH in absence of any
electric field where the top and bottom rows correspond to the LRH and SRH
magnetic helices, respectively.}
\label{tran}
\end{figure}

To unravel the basic mechanism of electron transport through any conducting
junction, it is always helpful to start with analyzing the nature of 
transmission function for the junction setup. The results shown in
Fig.~\ref{tran} describe the spin dependent transmission probabilities for
the LRH and SRH magnetic helices. For the time being we fix the 
external electric field to zero, and its effect will be considered in the
subsequent sections. Several notable features are observed from 
Fig.~\ref{tran} which we describe one by one as follows. A distinct feature
in the arrangements of resonant transmission peaks, associated with the
energy eigenvalues of the MH, can be observed in the short-range and 
long-range hopping helical geometries. For the SRH MH, the resonant peaks 
are almost uniformly arranged and their widths are nearly identical. This 
behavior 
is quite analogous to the conventional NNH model. On the other hand, for 
the other MH with LRH integrals, the resonant peaks are densely packed in
one side (left one), whereas they are widely separated in the other side
of the $T_{\sigma}$-$E$ spectrum (upper row of Fig.~\ref{tran}). This is 
\begin{figure}[ht]
{\centering\resizebox*{8cm}{3.5cm}{\includegraphics{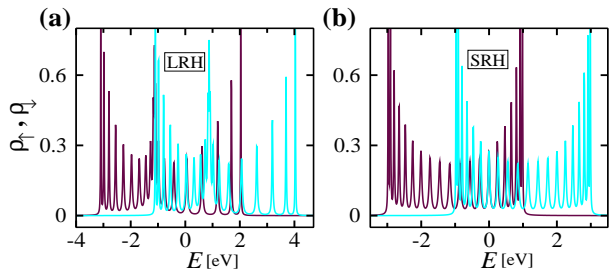}}\par}
\caption{(Color online). Density of states of up (maroon color) and down 
(cyan color) spin electrons for the identical magnetic helix systems as
taken in Fig.~\ref{tran}.}
\label{dos1}
\end{figure}
the generic feature of LRH model emphasizing the breaking of particle-hole
symmetry, unlike the NNH case where such symmetry is no longer violated.
The nature of short or long-range hopping of electrons can be visualized
by noting the distances between different neighbor sites. For the chosen
set of parameter values if we calculate the distances between first few 
neighbors then they will look like (in unit of $\AA$) for the LRH MH as:
$l_1=4.113$, $l_2=5.766$, $l_3=5.148$, $l_4=6.239$, $l_5=8.850$,  
$\dots$, while for the SRH one they are: $l_1=5.502$, $l_2=10.675$,
$l_3=15.242$, $l_4=19.033$, $l_5=22.023$, $\dots$, etc. As the neighboring
distances increase so rapidly for the SRH helix, electrons are no longer 
able to hop into far away sites, and the contributions essentially restricted
within the too few neighboring sites.

Now, due to the existence of spin dependent scattering term in the MH, 
as prescribed in Eq.~\ref{equ2}, a finite shift in energy scale between 
the up and down
spin transmission spectra takes place, associated with the two spin channels.
Thus, naturally when the Fermi energy is fixed anywhere within the region 
where only up or down spin transmission probability is finite, we can get 
pure up or down spin propagation through the junction yielding $100\%$
polarized spin current. This fact is quite well-known in literature in the
context of spin polarization considering a magnetic material. But, the 
interesting thing is that due to non-uniform distribution of $T_{\sigma}$-$E$
spectrum across its center for the case of LRH MH (upper row of 
Fig.~\ref{tran}), a significantly large spin polarization will be achieved 
even when the Fermi energy is placed within the overlap region between the 
two spin channels. This phenomenon, viz high degree of polarization, cannot 
be observed for the case of SRH helix, due to regular distribution of 
transmission functions where both up and down spin electrons contribute 
almost in equal amounts. The other key advantage of getting large spin 
polarization towards the band center is that experimentally it will be quite
\begin{figure}[ht]
{\centering\resizebox*{8cm}{7cm}{\includegraphics{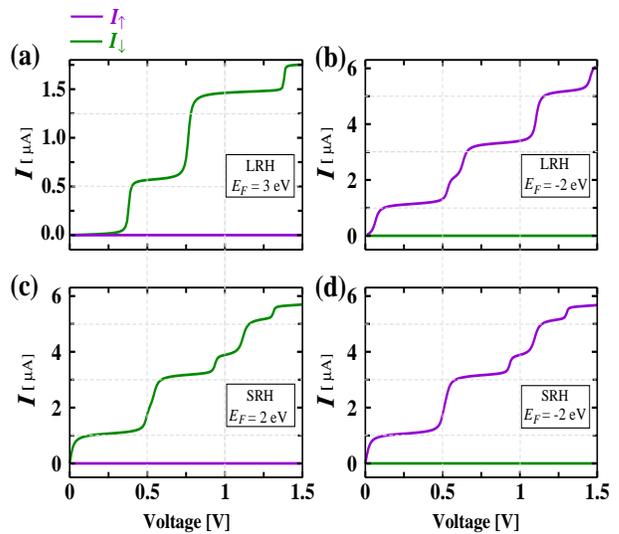}}\par}
\caption{(Color online). Current-voltage characteristics for the LRH and SRH
magnetic helix structures at some typical values of $E_F$, in the absence of
external electric field.}
\label{current}
\end{figure}
easier to place the Fermi energy and it can be tuned as well in these 
energy regions, rather than the placing of $E_F$ near the band edge. Thus,
the LRH MH is relatively superior than the SRH one, and more favorable 
justification for this statement can be found from our forthcoming analysis.

The origin of the finite shift of up and down spin transmission spectra
along with the overlap of these two spin bands, and the position of 
different transmission peaks in the LRH and SRH magnetic helices can be 
understood
in a more transparent way by noting the density of states (DOS) spectra
of different spin electrons for these magnetic helix structures. This is
due to the fact that the transmission function is directly related to
the energy spectrum and thus DOS. We compute DOS following the well-known 
relation $\rho_{\sigma}(E)=-(1/N\pi) \mbox{Im}[\mbox{Tr} 
(\mbox{\boldmath{$G$}}_{\sigma}^r)]$. Figure~\ref{dos1} shows the DOS
profiles for the up and down spin electrons for the identical magnetic
helix structures as considered in Fig.~\ref{tran}. From the spectra it
is noticed that a finite shift in energy scale takes place between the
two spin bands with a suitable overlap among them. At the same time the
alignments of different peaks along with their separations for the 
two different hopping cases are also nicely emerged. These features are
exactly evolved in the transmission spectra (Fig.~\ref{tran}).

The above nature of transmission function is directly reflected in 
current-voltage
characteristics. As illustrative examples, in Fig.~\ref{current} we show
the variation of spin dependent currents as a function of bias voltage $V$
at some typical Fermi energies. Here we also consider the zero-field case 
\begin{figure}[ht]
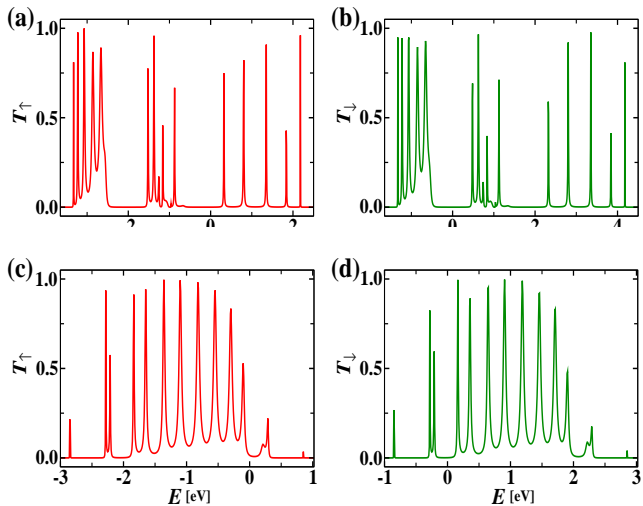

{\centering\resizebox*{4.2cm}{3.3cm}{\includegraphics{protranEfig1.eps}}
\resizebox*{4.2cm}{3.3cm}{\includegraphics{protranEfig2.eps}}
\resizebox*{4.2cm}{3.3cm}{\includegraphics{dnatranEfig1.eps}}
\resizebox*{4.2cm}{3.3cm}{\includegraphics{dnatranEfig2.eps}}\par}
\caption{(Color online). Up and down spin transmission probabilities in
presence of external gate voltage $V_g$ that produces the electric field
$E_g$ where the top and bottom rows correspond to the LRH and SRH helices,
respectively. Here we set $V_g=1\,$V.}
\label{tranE}
\end{figure}
\begin{figure}[ht]
{\centering\resizebox*{8.2cm}{7cm}{\includegraphics{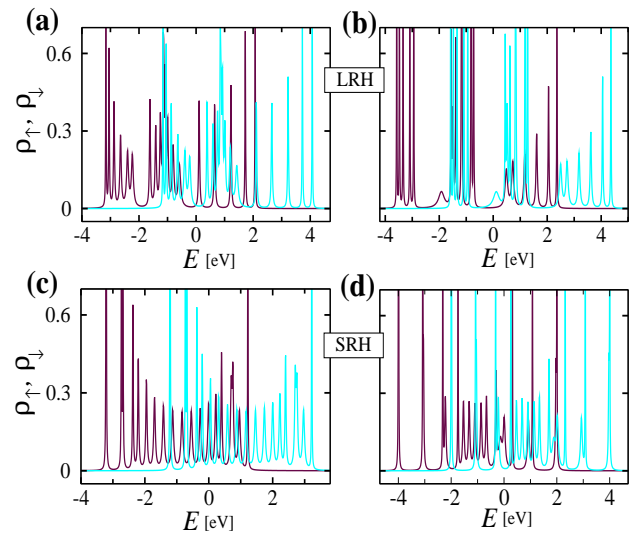}}\par}
\caption{(Color online). Density of states for up (maroon color) and 
down (cyan color) spin electrons in presence of external gate voltage
where the first column corresponds to $V_g = 0.5\,$V and $\beta=0$,
and the second column represents $V_g=1.5\,$V and $\beta=\pi/2$.}
\label{dos2}
\end{figure}
\begin{figure}[ht]
{\centering\resizebox*{8.2cm}{7cm}{\includegraphics{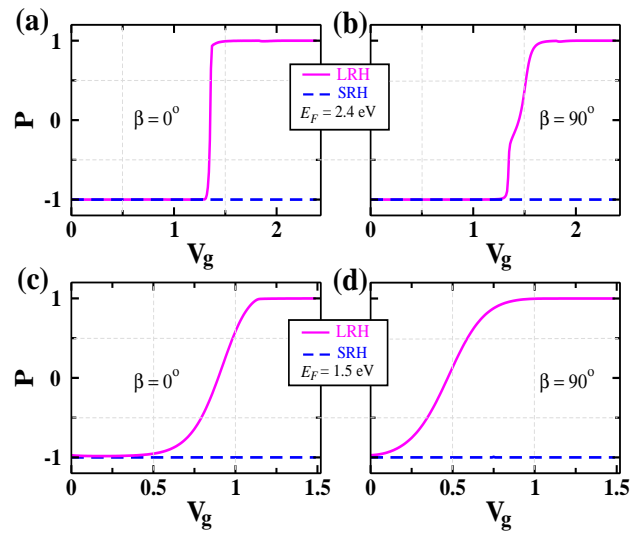}}\par}
\caption{(Color online). Spin polarization $P$ as a function
of gate voltage $V_g$ at some typical values of $\beta$ for two different
Fermi energies considering the bias voltage $V=0.2\,$V.}
\label{polvg}
\end{figure}
like Fig.~\ref{tran}, to visualize the reflectance of $T_{\sigma}$-$E$
spectra on $I_{\sigma}$-$V$ characteristics. The voltage window and the Fermi
energies are chosen selectively so that only one spin electrons can propagate
through the junction. Thus, for this entire bias window ($0<V<1.5\,$V), which
is reasonably large, only up or down spin electrons propagates yielding 
a $\pm 100\%$ spin polarized current. For the LRH MH, a distinct feature
in current steps is observed for the two different choices of Fermi energies,
$E_F=-2\,$eV and $E_F=3\,$eV. And the other notable thing is that the 
magnitudes of the currents are largely different, that can be visualized by
comparing the results shown by the green line in Fig.~\ref{current}(a) and
magenta line in Fig.~\ref{current}(b). These features are solely associated
with the transmission spectra (see the upper row of Fig.~\ref{tran}), as
the current is evaluated by the integration procedure of the transmission
function. On the other hand, for the SRH MH, the currents are almost
comparable to each other (green curve in Fig.~\ref{current}(c) and magenta
curve in Fig.~\ref{current}(d)), following the uniform distribution of 
transmission peaks.

From the zero-field spin dependent transmission probabilities 
(Fig.~\ref{tran}) and currents (Fig.~\ref{current}) we already get some 
suitable hints
about the interplay between the two kinds of hopping integrals on spin 
selective transmission. Now, in the remaining parts we include the 
effects of external electric field to explore several more interesting 
features in the magnetic helix geometry. Like above, here also we start with
the variation of transmission function with energy $E$. The results are shown 
in Fig.~\ref{tranE}. In the presence of electric field, the system behaves 
like a correlated disordered one~\cite{cor1,cor2} and therefore the 
transmission spectrum is no longer symmetric even for the short-range 
hopping magnetic helix. The key feature is that, the transmission spectrum 
is {\em gapped}~\cite{cor1,cor2,fra1,fra2} associated with the energy
spectrum of the MH, and it is more prominent when we consider longer-range 
hopping of electrons. Due to this gapped spectrum, there is a finite 
probability to have non-zero transmission of one spin electrons at multiple 
energy zones, even not far away from the band center, while completely 
vanishing transmission for the other spin electrons.
Under this situation we can get $100\%$ polarized spin current. The
width of these gaps and at the same time swapping of contributing channel 
can also be tuned with the help of $V_g$ as well as phase factor $\beta$
that we confirm through our exhaustive calculations. 

The gapped nature of the transmission functions and their possible tuning
by means of $V_g$ and $\beta$ can be understood easily by noting the 
density of states (DOS) profiles under different input conditions of these
\begin{figure}[ht]
{\centering \resizebox*{7cm}{5.3cm}{\includegraphics{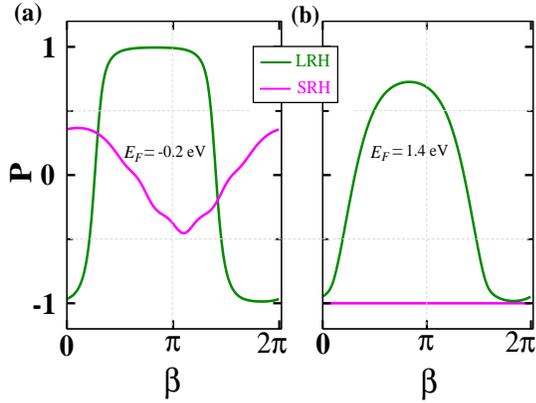}}\par}
\caption{(Color online). Dependence of spin polarization $P$ with phase factor
$\beta$ at two different Fermi energies. Here we fix $V_g=0.5\,$V and
$V=0.2\,$V.}
\label{beta}
\end{figure}
physical parameters, as the transmission spectrum is directly related to
the energy spectrum and thus the DOS. 
Figure~\ref{dos2} displays the DOS spectra for up and down spin electrons 
at some typical values of $V_g$ and $\beta$. Prominent bands, separated by
sharp gaps, are obtained for the LRH MH, whereas the effects are 
\begin{figure}[ht]
{\centering\resizebox*{6cm}{7cm}{\includegraphics{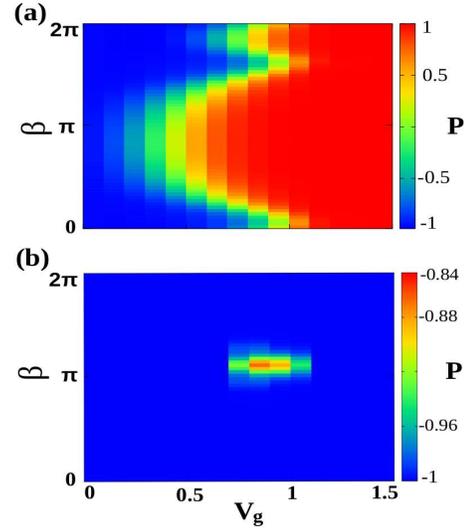}}\par}
\caption{(Color online). Simultaneous variation of $P$ with $V_g$ and $\beta$
for the (a) LRH and (b) SRH helices, considering $V=0.2\,$V and $E_F=1.5\,$eV.}
\label{density}
\end{figure}
comparatively less for the SRH helix. The shifting of different peaks with
the inclusion of $\beta$ is clearly visible. The transmission functions 
corroborate these behaviors what are shown in Fig.~\ref{tranE}.

These facts can be validated further from the results presented in 
Fig.~\ref{polvg}, where the dependence of spin polarization $P$ is shown
as a function of gate voltage for some typical values of phase factor 
$\beta$ at two Fermi energies. The bias voltage $V$ is fixed at $0.2\,$V.
From all the spectra it is clearly seen that much higher spin polarization
can be obtained, and in some cases it reaches to $100\%$, by varying 
$V_g$. The notable feature is that a perfect phase reversal of spin 
polarization i.e., $-100\%$ to $+100\%$ can be done by regulating the gate
voltage. This is a clear signature of externally tuning selective spin 
transmission using a magnetic helix structure.

The above discussion raises an obvious question that as the change of site
energy, due to the variation of $V_g$, leads to a significant change in 
\begin{figure}[ht]
{\centering \resizebox*{8cm}{7cm}{\includegraphics{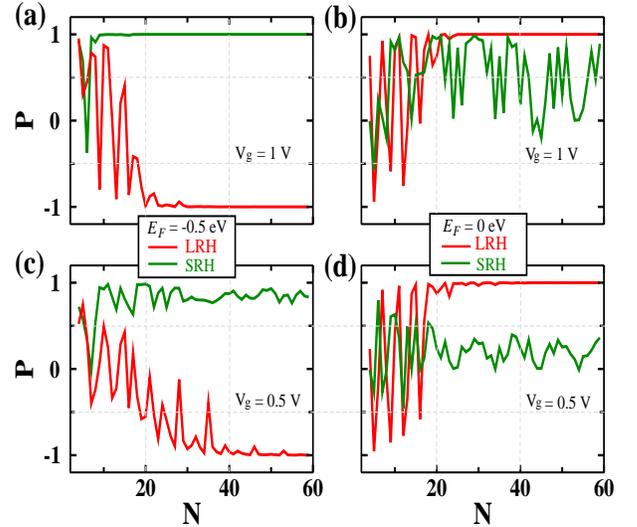}}\par}
\caption{(Color online). Effect of system size $N$ on spin polarization
$P$. Here we choose $V=0.2\,$V.}
\label{size}
\end{figure}
transmission spectra, and thus, spin polarization, can we also expect any 
non-trivial signature of selective spin transmission by tuning the phase
factor $\beta$ for a constant $V_g$, as it is directly involved in site 
energy (see Eq.~\ref{equ5}). To explore it, let us focus on the spectra
given in Fig.~\ref{beta}, where the variation of $P$ as a function of $\beta$
is shown for two typical Fermi energies. The results are computed setting
$V_g=0.5\,$V and $V=0.2\,$V. For the SRH MH a minor variation of $P$ is 
noticed with $\beta$, whereas a significantly large change in $P$ is possible 
for the LRH MH, which again proves the superiority of LRH MH compared to
the other one. The underlying mechanism is same as noted earlier that the
contributing spin channels are modified with the modulation of site energy.
Thus, for a fixed gate voltage, selective spin polarization can be achieved
simply by changing the orientation of incident electric field. This is a more
suitable way of tuning spin transport compared to the other proposals where
usually magnetic field plays this role.

To test the sensitivity of $P$ on both the gate voltage and phase
factor, in Fig.~\ref{density} we plot the simultaneous variation of $P$ with
$V_g$ and $\beta$, considering $V=0.2\,$V and $E_F=1.5\,$eV. A drastic change 
is clearly visible between the LRH and SRH helices. The spin polarization $P$
is almost insensitive for a wide range of both $V_g$ and $\beta$ in the 
case of SRH MH, where it varies very close to $-1$. On the other hand, for 
the other situation, a complete phase reversal of $P$ is possible, and each 
of these two phases ($P=+1$ or $P=-1$) persists over a wide range of 
parameter values, which thus gives us a confidence to examine the results 
in laboratory.

In this context it is also relevant to examine the role of system size $N$. 
The results are presented in Fig.~\ref{size} for two specific values of $V_g$
setting the voltage bias at $0.2\,$V. In the case of long-range hopping 
magnetic helix, large oscillation is observed for smaller $N$, and it is 
gradually dying out with $N$. While, a finite oscillation with almost 
constant amplitude is exhibited for the case of short-range hopping helix. 
The strong modulation of spin polarization with helix size $N$ appears 
due to the quantum interference of electronic waves, and this effect is
again directly related to the incomplete helix nature with the change of 
$N$.

\subsection{CISS effect}

It is well known that a simple ferromagnetic material exhibits spin 
filtration effect. This is quite trivial and extensively studied in 
\begin{figure}[ht]
{\centering \resizebox*{8.2cm}{4.3cm}{\includegraphics{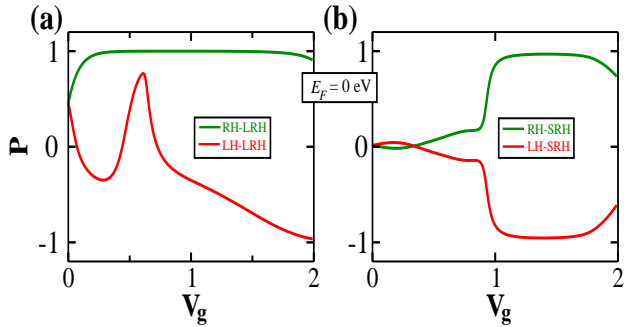}}\par}
\caption{(Color online). Spin polarization coefficient $P$ as function of
gate voltage $V_g$ for the left- and right-handed helical geometries.
We set $N=19$ and $21$, respectively, for the LRH and SRH magnetic helices, 
to have the full turns. The other physical parameters are: bias voltage 
$V=0.2\,$V and the phase factor $\beta=\pi/2$.} 
\label{chrl1}
\end{figure}
literature. But, the magnetic helix geometry which we consider here in 
our work is completely a new configuration, as it exhibits strong CISS 
effect, analogous to what was discussed by some pioneering groups in their 
works~\cite{ghl,ciss1,ciss2,ciss3,ciss4,ciss5}. To the best of our knowledge, 
no one has explored the CISS effect in any MH geometry so far. 

To reveal the helicity effect on spin polarization, let us look into the
spectra given in Fig.~\ref{chrl1} where we show the variation of $P$
on gate voltage $V_g$, for the left-handed (LH) and right-handed (RH)
helical geometries. The results are computed for exactly two opposite 
helcities, keeping all the other parameters unchanged, where (a) and (b)
correspond to the LRH and SRH helices, respectively. To get the magnetic
helix with full number of turns (which is extremely important to examine
the CISS effect and compare the results between two different handed helical
geometries), here we take $N=19$ for the long-range hopping MH, while it
($N$) becomes $21$ for the short-range MH. The spin polarization 
changes drastically with the handedness, as clearly seen by comparing 
the two colored curves. It becomes more prominent for the SRH model 
compared to the LRH one. Almost a complete phase reversal of $P$ takes
place upon the alteration of the helical sense. The nature of 
polarization with the change of handedness is solely associated with the
contributing up and down spin channels within the selected energy window
around $E_F$, associated with the bias voltage. For the short-range 
helix, almost perfect swapping takes place between the up and down spin 
transmission probabilities upon the inversion of handedness, which results
nearly full inversion of spin polarization. On the other hand, for the
long-range hopping helix, the transmission probabilities are largely
asymmetric and perfect swapping is no longer possible, and thus, we do 
not get the complete phase reversal of $P$ for the opposite handed helix 
geometry, like what we see in the case of short-range helix system.

Keeping the handedness (right- or left-handed) of the MH fixed, we can 
\begin{figure}[ht]
{\centering \resizebox*{8.2cm}{4.3cm}{\includegraphics{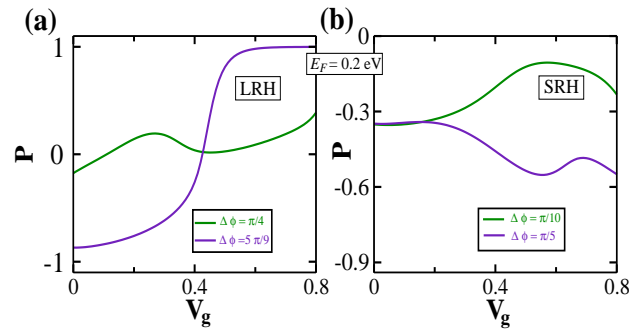}}\par}
\caption{(Color online). Dependence of spin polarization $P$ on $V_g$
at two different values of $\Delta\phi$, where (a) $N=19$ and (b) $N=21$. 
Here we choose $V=0.2\,$V and $\beta=\pi/2$.}
\label{chrl2}
\end{figure}
further investigate the role of helicity by changing the geometrical
conformation associated with the twisting. With the reduction of the
twisting angle $\Delta \phi$, the helical shape gradually transforms
into the linear-like geometry, and eventually becomes a magnetic chain
when $\Delta \phi=0$. This conformational effect should be directly
reflected in spin selective electron transmission. This is exactly what
we see from Fig.~\ref{chrl2}, where two typical values of $\Delta \phi$ 
are taken into account. Both for the LRH and SRH magnetic helices, we find 
a strong dependence of spin polarization on the twisting angle. For a fixed 
hopping helix, be it a LRH or SRH one, the degree of polarization sharply 
decreases with reducing the twisting angle, and the gate voltage $V_g$ does 
not have such an important role under this situation. So, undoubtedly, 
helicity has an important role in spin polarization.

\subsection{Effect of helical dynamics on spin selective transmission}

As already noted, for the complete scenario and to have the model more 
\begin{figure}[ht]
{\centering \resizebox*{8cm}{7cm}{\includegraphics{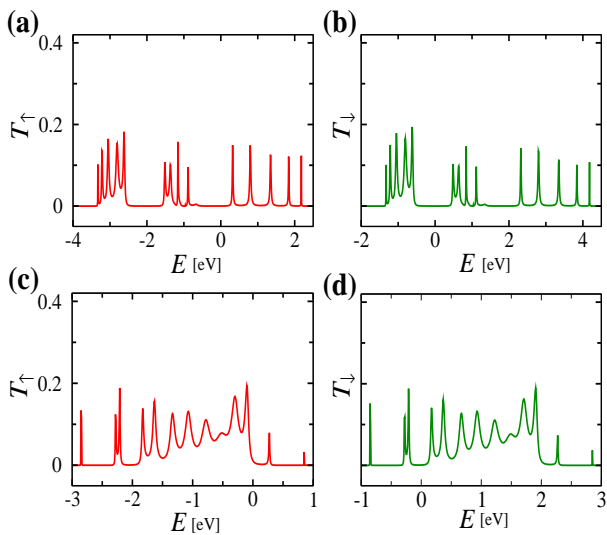}}\par}
\caption{(Color online). Up and down spin transmission probabilities in
presence of external gate voltage $V_g$, where the top and bottom rows
correspond to the LRH and SRH helices, respectively. Here we set $V_g=1\,$V,
$\beta=0$ and the decay time $\tau=1\,$fs.}
\label{dyna1}
\end{figure}
\begin{figure}[ht]
{\centering\resizebox*{8.2cm}{7cm}{\includegraphics{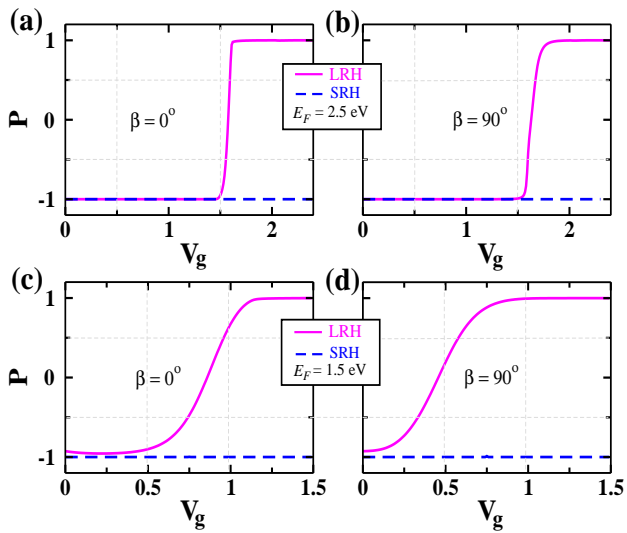}}}
\caption{(Color online). $P$-$V_g$ characteristics at some typical values
of $\beta$ for two different Fermi energies. Here we set $V=0.2\,$V and
$\tau=1\,$fs.}
\label{dyna2}
\end{figure}
realistic we need to include the effect of helical dynamics. In this
sub-section we critically examine this issue, where the dynamical effect 
is included following the prescription given by Ratner and 
co-workers~\cite{hd2}.

Let us begin with Fig.~\ref{dyna1}, where spin dependent transmission 
probabilities are shown for the LRH helix (upper panel) and the SRH helix 
(lower panel), considering the decay time 
$\tau=1\,$fs, keeping all the other parameters unchanged as considered in
Fig.~\ref{tranE}, to have a better comparison of the spectra obtained for
the static (Fig.~\ref{tranE}) and dynamic (Fig.~\ref{dyna1}) cases. At a 
\begin{figure}[ht]
{\centering \resizebox*{8.3cm}{5.3cm}{\includegraphics{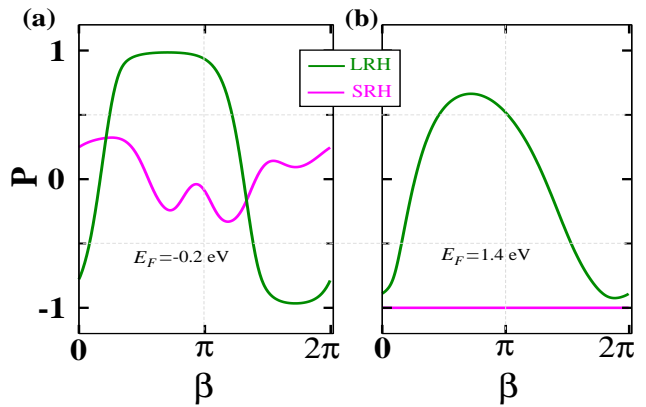}}\par}
\caption{(Color online). Dependence of spin polarization $P$ with phase factor
$\beta$ for the two helical geometries. Here we fix $V_g=0.5\,$V, $V=0.2\,$V
and $\tau=1\,$fs.}
\label{dyna3}
\end{figure}
first glance we see that the {\em gapped} nature of the spectra remains
almost identical in the presence of the dynamical effect, like what we 
notice in the absence of this effect (see Fig.~\ref{tranE}). The significant
change occurs in transmission amplitudes at the resonances. The amplitude
reduces drastically with the helical dynamics. These features can be 
explained as follows. Due to the addition of a complex part at the acceptor
\begin{figure}[ht]
{\centering \resizebox*{8.2cm}{3.5cm}{\includegraphics{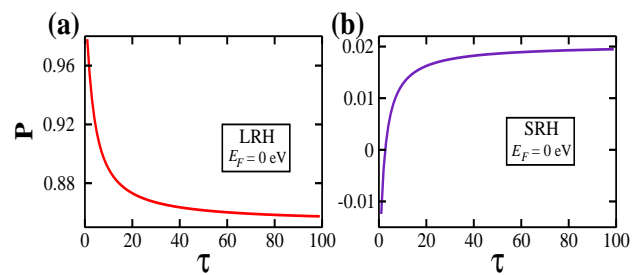}}\par}
\caption{(Color online). Polarization $P$ as a function of decay time $\tau$
(in unit of fs) for the LRH and SRH helices. Here we set $V_g=0.5\,$V,
$V=0.2\,$V and $\beta=\pi/2$.}
\label{dyna4}
\end{figure}
site (viz, the site of the MH where the drain electrode is coupled), the 
effective coupling gets reduced. This coupling significantly influences
the electron transfer across the nanojunction, yielding lesser transmission 
amplitudes.

Though the transmission amplitudes get reasonably hampered due to the
dynamical effect, it does not have any considerable influence on spin
polarization. It is clearly reflected from the spectra presented in
Fig.~\ref{dyna2}, where the results are computed for $\tau=1\,$fs, setting
all the other physical parameters constant as considered in Fig.~\ref{polvg}.
The underlying physics is that, the spin polarization depends on the ratios
of two spin dependent currents. Thus, the reduction of both these two spin
currents results almost same degree of spin polarization like what we get
in the absence of dynamical effect. It essentially gives a strong confidence
that the magnetic helix can be utilized for efficient spin filtration as
the spin polarization has not been largely perturbed even in the presence 
of dynamical effects.

This claim can be justified further by looking into the spectra given in
Fig.~\ref{dyna3}, where the dependence of spin polarization with phase 
factor $\beta$ is shown, like what are presented earlier in the absence of
helical dynamics (see Fig.~\ref{beta}). Almost identical scenario is
noticed in both these two cases i.e., in the absence and presence of 
helical dynamics.

Finally, to have the more comprehensive picture of the effect of helical
dynamics, in Fig.~\ref{dyna4} we show the dependence of spin polarization
on the decay time $\tau$, by varying it in a wide range setting the Fermi
energy at zero. Both for the LRH and SRH magnetic helices, we find that 
the change of $P$ is too small, and remains almost constant for a wide 
range of $\tau$. This behavior strengthens the robustness of our analysis 
in the magnetic helix geometry.

\section{Summary and Outlook}

To conclude, in the present work first time we have addressed the issues of
spin selective transmission through a {\em single-stranded magnetic helix},
and thoroughly discussed the interplay between the short- and long-range
hopping integrals, CISS effect and the role of helical dynamics. We have 
also explored how to regulate transport properties {\em externally} by 
applying a suitable electric field and changing its orientation, instead 
of applying a magnetic field as conventionally considered to regulate 
electron transfer through magnetic materials. All the characteristic
features have been studied using the standard Green's function formalism 
within the tight-binding framework. The key findings and the important 
aspects are summarized as follows. \\
$\bullet$ High degree of spin polarization can be achieved in a magnetic 
helix, and the performance becomes more superior in the long-range hopping
helix rather than the short-range hopping helical geometry. \\
$\bullet$ A single-stranded MH is capable to yield spin selective transmission,
and most importantly, we do not need to include the effects of environmental
dephasing. \\
$\bullet$ The degree of spin polarization along with its phase (positive or
negative) can be regulated externally by applying an electric field. Thus, 
our proposal can be utilized to design externally controlled selective
spin transmission. \\
$\bullet$ The proposed magnetic helix structure exhibits pronounced CISS 
effect, similar to what is observed in chiral molecules like DNA and 
others. \\
$\bullet$ The degree of spin polarization has not been perturbed reasonably
even in the presence of helical dynamics. This phenomenon certainly 
strengthens the robustness of our analysis. \\
$\bullet$ All the results are valid for a wide range of physical parameters,
which prove the robustness of our analysis, and give us a confidence to
examine our results in a suitable laboratory setup. \\
$\bullet$ Our prescription can be implemented in any other magnetic helical 
geometry having LRH or SRH integrals.

\section{Acknowledgments}

SS is grateful to CSIR, India (File number: 09/093(0183)/2017-EMR-I) for 
providing her research fellowship. The research of SKM is supported by 
DST-SERB, India, under Grant number EMR/2017/000504. SKM thanks very much
Prof. Shreekantha Sil and Mrs. Moumita Patra for useful discussions. We
thank all the reviewers for their valuable comments and suggestions to
enhance the quality of the paper.


\begin{thebibliography}{99}

\bibitem{wolf} S. A. Wolf, D. D. Awschalom, R. A. Buhrman, J. M. Daughton, 
S. von Moln\'{a}r, M. L. Roukes, A. Y. Chtchelkanova, and D. M. Treger, 
Science \textbf{294}, 1488 (2001).

\bibitem{s1} D. E. Nikonov, G. I. Bourianoff, and P. A. Gargini, 
J. Supercond. Novel Magn. \textbf{19}, 497 (2006).

\bibitem{s2} M. Johnson and R. H. Silsbee, Phys. Rev. Lett. \textbf{55},
1790 (1985).

\bibitem{s3} M. N. Baibich, J. M. Broto, A. Fert, F. Nguyen Van Dau, 
F. Petroff, P. Etienne, G. Creuzet, A. Friederich, and J. Chazelas, 
Phys. Rev. Lett. \textbf{61}, 2472 (1988).

\bibitem{s4} I. Zuti\'{c}, J. Fabian, and S. Das Sarma, Rev. Mod. Phys. 
\textbf{76}, 323 (2004).

\bibitem{s5} S. Datta and B. Das, Appl. Phys. Lett. \textbf{56}, 665 (1990).

\bibitem{s6} J. P. Lu, J. B. Yau, S. P. Shukla, M. Shayegan, L. Wissinger, 
U. R\"{o}ssler, and R. Winkler, Phys. Rev. Lett. \textbf{81}, 1282 (1998).

\bibitem{xie} P. Zhang, Q. K. Xue, and X. C. Xie, Phys. Rev. Lett.
\textbf{91}, 196602 (2003).

\bibitem{wang} W. Long, Q. F. Sun, H. Guo, and J. Wang, Appl. Phys. Lett.
\textbf{83}, 1397 (2003).

\bibitem{int1} T. P. Pareek, Phys. Rev. Lett. \textbf{92}, 076601 (2004).

\bibitem{int2} Q. F. Sun and X. C. Xie, Phys. Rev. B \textbf{73},
235301 (2006).

\bibitem{int3} F. Chi, J. Zheng, and L. L. Sun, Appl. Phys. Lett.
\textbf{92}, 172104 (2008).

\bibitem{rashba} Y. A. Bychkov and E. I. Rashba, Pis'Ma Zh. Eksp. Teor. 
Fiz. \textbf{39}, 66 (1984); [JETP Lett. \textbf{39}, 78 (1984)].

\bibitem{dressel} G. Dresselhaus, Phys. Rev. \textbf{100}, 580 (1955).

\bibitem{winkler} R. Winkler, {\em Spin-orbit coupling effects in
two-dimensional electron and hole Systems}, Springer Tracts in Modern Physics,
Springer, New York, Vol. 191, (2003).

\bibitem{ex1} J. Nitta, T. Akazaki, H. Takayanagi, and T. Enoki,
Phys. Rev. Lett. \textbf{78}, 1335 (1997).

\bibitem{ex2} J. P. Heida, B. J. van Wees, J. J. Kuipers, T. M. Klapwijk,
and G. Borghs, Phys. Rev. B \textbf{57}, 11911 (1998).

\bibitem{ex3} D. Grundler, Phys. Rev. Lett. \textbf{84}, 6074 (2000).

\bibitem{ex4} T. Matsuyama, R. K\"{u}rsten, C. Meiner, and U. Merkt,
Phys. Rev. B \textbf{61}, 15588 (2000).

\bibitem{pr1} A. A. Kislev and K. W. Kim, J. App. Phys. \textbf{94}, 4001
(2003).

\bibitem{pr2} I. A. Shelykh, N. G. Galkin, and N. T. Bagraev,
Phys. Rev. B \textbf{72}, 235316 (2005).

\bibitem{pr3} P. F\"{o}ldi, O. K\'{a}lm\'{a}n, M. G. Benedict, and
F. M. Peeters, Phys. Rev. B \textbf{73}, 155325 (2006).

\bibitem{pr4} G. Cohen, O. Hod, and E. Rabani, Phys. Rev. B \textbf{76},
235120 (2007).

\bibitem{pr5} S. Ganguly, S. Basu, and S. K. Maiti, Europhys. Lett. 
\textbf{124}, 17005 (2018). 

\bibitem{pr6} S. Ganguly, S. Basu, and S. K. Maiti, Superlattices 
Microstruct. \textbf{120}, 650 (2018). 

\bibitem{pr7} W. J. M. Naber, S. Faez, and W. G. van der Wiel, J. Phys. D: 
Appl. Phys. \textbf{40}, R205 (2007).

\bibitem{pr8} I. Bergenti, V. Dediu, M. Prezioso, and A. Riminucci, 
Philos. Trans. R. Soc. London, Ser. A \textbf{369}, 3054 (2011).

\bibitem{ghl} B. G\"{o}hler, V. Hamelbeck, T. Z. Markus, M. Kettner, 
G. F. Hanne, Z. Vager, R. Naaman, and H. Zacharias, Science \textbf{331}, 
894 (2011). 

\bibitem{ciss1} Z. Xie, T. Z. Markus, S. R. Cohen, Z. Vager, R. Gutierrez,
and R. Naaman, Nano Lett. \textbf{11}, 4652 (2011).

\bibitem{ciss2} R. Naaman and D. H. Waldeck, J. Phys. Chem. Lett. 
\textbf{3}, 2178 (2012).

\bibitem{ciss3} K. Senthil Kumar, N. Kantor-Uriel, S. P. Mathew, R. Guliamov,
and R. Naaman, Phys. Chem. Chem. Phys. \textbf{15}, 18357 (2013).

\bibitem{ciss4} R. Guti\'{e}rrez, E. D\'{i}az, R. Naaman, 
and G. Cuniberti, Phys. Rev. B \textbf{85}, 081404(R) (2012).

\bibitem{ciss5} E. Medina, F. L\'{o}pez, M. A. Ratner, 
and V. Mujica, Europhys. Lett. \textbf{99}, 17006 (2012).

\bibitem{qfs1} A.-M. Guo and Q.-F. Sun, Phys. Rev. Lett. \textbf{108}, 
218102 (2012).

\bibitem{qfs2} A.-M. Guo and Q.-F. Sun, Proc. Natl. Acad. Sci. U.S.A. 
\textbf{111}, 11658 (2014).

\bibitem{qfs3} T.-R. Pan, A.-M. Guo, and Q.-F. Sun, Phys. Rev. B \textbf{92}, 
115418 (2015).

\bibitem{smSO} F. Kuemmeth, S. Ilani, D. C. Ralph, and P. L. McEuen, 
Nature (London) \textbf{452}, 448 (2008).

\bibitem{gc1} K.-H. Yoo, D. H. Ha, J.-O. Lee, J. W. Park, J. Kim, 
J. J. Kim, H.-Y. Lee, T. Kawai, and H. Y. Choi, Phys. Rev. Lett. \textbf{87}, 
198102 (2001).

\bibitem{gc2} G. Maruccio, A. Biasco, P. Visconti, A. Bramanti, P. P. Pompa, 
F. Calabi, R. Cingolani, R. Rinaldi, S. Corni, R. Di Felice, E. Molinari, 
M. P. Verbeet, and G. W. Canters, Adv. Mater. \textbf{17}, 816 (2005). 

\bibitem{gc3} K. Bradley, M. Briman, A. Star, and G. Gr\"{u}ner, 
Nano Lett. \textbf{4}, 253 (2004).

\bibitem{gc4} A. V. Malyshev, Phys. Rev. Lett. \textbf{98}, 096801 (2007).

\bibitem{gc5} A.-M. Guo and Q.-F. Sun, Phys. Rev. B \textbf{86}, 035424 
(2012).

\bibitem{hd1} R. Guti\'{e}rrez, R. Caetano, P. B. Woiczikowski,
T. Kubar, M. Elstner, and G. Cuniberti, New. J. Phys. \textbf{12}, 023022
(2010).

\bibitem{hd2} F. C. Grozema, Y. A. Berlin, L. D. A. Siebbeles,
and M. A. Ratner, J. Phys. Chem. B \textbf{114}, 14564 (2010).

\bibitem{hd3} F. C. Grozema, S. Tonzani, Y. A. Berlin,
G. C. Schatz, L. D. A. Siebbeles, and M. A. Ratner, J. Am. Chem. Soc.
\textbf{130}, 5157 (2008).

\bibitem{hd4} K. Senthilkumar, F. C. Grozema, C. F. Guerra,
F. M. Bickelhaupt, F. D. Lewis, Y. A. Berlin, M. A. Ratner, and 
L. D. A. Siebbeles, J. Am. Chem. Soc. \textbf{127}, 14894 (2005).

\bibitem{hd5} F. C. Grozema, Y. A. Berlin, and L. D. A. 
Siebbeles, J. Am. Chem. Soc. \textbf{122}, 10903 (2000).

\bibitem{hd6} A. Troisi and G. Orlandi, J. Phys. Chem. B 
\textbf{106}, 2093 (2002).

\bibitem{hd7} R. N. Barnett, C. L. Cleveland, A. Joy, 
U. Landman, and G. B. Schuster, Science \textbf{294}, 567 (2001).

\bibitem{spsp} Y.-H.Su, S.-H. Chen, C. D. Hu, and 
C.-R. Chang, J. Phys. D: Appl. Phys. \textbf{49}, 015305 (2016).

\bibitem{mh1} A. A. Shokri, M. Mardaani, and K. Esfarjani, Physica E 
\textbf{27}, 325 (2005).

\bibitem{mh2} A. A. Shokri and M. Mardaani, Solid State Commun. \textbf{137}, 
53 (2006).

\bibitem{mh3} M. Dey, S. K. Maiti, and S. N. Karmakar, Phys. Lett. A 
\textbf{374}, 1522 (2010).

\bibitem{mh4} M. Patra and S. K. Maiti, Sci. Rep. \textbf{7}, 14313 (2017).

\bibitem{efl} A.-M. Guo and Q.-F. Sun, Phys. Rev. B \textbf{95}, 155411
(2017).

\bibitem{gf1} S. Datta, {\it Electronic Transport in Mesoscopic Systems},
Cambridge University Press, Cambridge (1997).

\bibitem{gf2} D. S. Fisher and P. A. Lee, Phys. Rev. B {\bf 23}, 6851 (1981).

\bibitem{gf3} M. Dey, S. K. Maiti, and S. N. Karmakar, Org. Electron.
\textbf{12}, 1017 (2011).

\bibitem{pola1} D. Rai and M. Galperin, Phys. Rev. B \textbf{86}, 045420 
(2012).

\bibitem{rps} R. G. Endres, D. L. Cox, and R. R. P. Singh, Rev. Mod. 
Phys. \textbf{76}, 195 (2004).

\bibitem{cor1} S. Ganeshan, K. San, and S. Das Sarma, Phys. Rev. Lett.
\textbf{110}, 180403 (2013).

\bibitem{cor2} Y. E. Kraus, Y. Lahini, Z. Ringel, M. Verbin, and 
O. Zilberberg, Phys. Rev. Lett. \textbf{109}, 106402 (2012).

\bibitem{fra1} M. Kohmoto, B. Sutherland, and C. Tang, Phys. Rev. B 
\textbf{35}, 1020 (1987).

\bibitem{fra2} G. J. Jin, Z. D. Wang, A. Hu, and S. S. Jiang, 
Phys. Rev. B \textbf{55}, 9302 (1997).

\end{thebibliography}
\end{document}